# Quasi van der Waals Epitaxial Growth of GaAsSb Nanowires on Graphitic Substrate for Photonic Applications

Dingding Ren[†], Tron A. Nilsen[†], Julie S. Nilsen[‡], Lyubomir Ahtapodov[†], Anjan Mukherjee[†], Yang Li[†], Antonius T. J. van Helvoort[‡], Helge Weman[†] and Bjørn-Ove Fimland[†*]

[†] Department of Electronic Systems and [‡] Department of Physics, Norwegian University of Science and Technology (NTNU), NO-7491, Trondheim, Norway

[*]Correspondence should be addressed to: bjorn.fimland@ntnu.no

**ABSTRACT**. III-V semiconductor nanowires are considered promising building blocks for advanced photonic devices. One of the key advantages is that the lattice mismatch can easily be accommodated in 1D structures, resulting in superior heteroepitaxial quality compared to thin films. However, few reports break the limitation of using bulk crystalline materials as substrates for epitaxial growth of high-quality photonic 1D components, making monolithic integration of III-V components on arbitrary substrates challenging. In this work, we show that the growth of self-catalyzed GaAsSb nanowires on graphitic substrates can be promoted by creating step edges of monolayer thickness on kish graphite before the growth. By further alternating the deposition sequence of the group-III element Al and the group-V elements As and Sb, it was found that triangular crystallites form when Al is deposited first. This indicates that the surface binding energy between the graphitic surface and the III-V nucleus profoundly influences the epitaxial growth of III-V materials on graphitic surfaces. Using the optimized growth recipe with an AlAsSb buffer nuclei, vertical [111]-oriented GaAsSb/GaAs nanowires with GaAsSb-based multiple axial superlattices were grown on exfoliated graphite, which was attached to a (001) AlAs/GaAs distributed Bragg reflector (DBR) using the simple Scotch tape method. Fabry–Pérot resonance modes were observed under optical excitation at room temperature, indicating a successful monolithic integration with optical feedback from the DBR system. These results demonstrate the great potential for flexible integration of high-efficiency III-V nanowire photonic devices on arbitrary photonic platforms using a 2D material buffer layer, e.g., graphene, without breaking the orientation registry.

**MAIN TEXT.** Epitaxially grown III-V semiconductors on graphene is a promising alternative route for high-performance optoelectronic devices. Such devices can benefit from the unique properties of graphene as a robust transparent electrode due to its high carrier mobility[1], excellent mechanical properties[2], fast heat dissipation[3], and nearly full-spectrum transparency[4]. However, unlike traditional III-V or Si substrates, perfect graphene or graphitic surfaces are composed of thoroughly hybridized $sp^2$-bonded carbon atoms. These are chemically inert due to a lack of dangling bonds and it is therefore challenging to grow III-V semiconductors on them. Therefore, several efforts have been made to utilize remote epitaxy via the quasi van der Waals (qvdW) interaction[5] or the nanoscale pinhole-opening on graphene, thru-hole epitaxy[6,7], to grow III-V thin films on bulk crystalline substrates with graphene as an intermediate layer. However, for monolithic integration of III-V components on arbitrary substrates, there is a great demand to break the limitation of using lattice-matched bulk crystalline materials as substrates for epitaxial growth of high-quality photonic components. Multi-layer graphene (MLG) could provide a universal platform to study the qvdW epitaxial growth of III-V components on arbitrary substrates.[8–10] Firstly, MLG can be transferred to any substrate using the Scotch tape method or the PMMA-assisted transfer method.[11,12] Secondly, MLG is a few atomic layers in thickness, which is crucial because it keeps the good optoelectronic properties of single-layer graphene, like broadband optical transparency. More importantly, MLG is thick enough to exclude substrate interaction like remote epitaxy or thru-hole epitaxy. Although there are both practical demands and scientific interests, it is unclear how to grow high-quality III-V semiconductors on MLG based on qvdW epitaxy.

Due to the nanoscale footprint and efficient termination of threading dislocations on sidewalls, one-dimensional (1D) or quasi-1D semiconductor nanowires possess a unique geometry that does not require the strict lattice matching[13,14] or a single-crystalline substrate to grow from[15,16]. InAs[17], GaN[18] and ZnO[19] binary nanowires, as well as InAsSb[20] and InGaAs[21] ternary nanowires, have been successfully grown on graphene or graphitic surfaces. These materials share the same commonality that their lattice constants in the growth direction of [111] for zinc blende (ZB) or [0001] for wurtzite (WZ) are matched or nearly matched to that of graphene (either 3.26 Å or 6.1 Å, respectively)[9]. Between these lattice parameters are the materials with bandgaps in the ultraviolet to near-infrared regions, which are essential for solar cell, light emitting diode and photodetector applications. Although GaAs nanowires can be grown on graphitic surfaces using both molecular beam epitaxy (MBE) and metalorganic vapor-phase epitaxy (MOVPE), the vertical nanowire density is much lower than on conventional III-V substrates, and it remains challenging to grow III-V nanowires on graphitic surfaces.[9,22]

Sb-containing III-V ternary nanowires have attracted much attention due to their tunable bandgap from 0.7 eV to 1.4 eV for optical communication, solar energy harvesting and infrared sensing.[23–26] The growth of GaAsSb nanowires is commonly achieved via MBE and MOVPE. GaAsSb nanowires have been used for singe nanowire or array-based high-performance photodetectors[24,27], and GaAsSb-based nanowire superlattices show single-mode lasing in the near-infrared with a record-low threshold at room temperature[28]. In addition, incorporating Sb during III-V nanowire growth brings new perspectives in understanding the interplay among supersaturation[29], surface energy[30], and diffusion properties of III-V adatoms[31] with an improved vertical yield of nanowires[16], which are crucial for future quantum engineering of nanowires with atomic precision.[26,32] Lastly, GaAsSb nanowires show promise of epitaxial integration with graphene[33].

In this work, we systematically explore the nucleation window of III-V materials on graphitic substrates by creating step edges on graphite surface or by using atomic layer epitaxy (ALE) of an AlAsSb buffer. It was found that a higher adsorption energy is beneficial for nucleation formation and vertical nanowire growth. By using a simple graphene Scotch exfoliation transfer method and optimized nucleation conditions, vertical [111]-oriented GaAsSb/GaAs nanowires with GaAsSb-based multiple axial superlattices have been successfully grown on a (001) AlAs/GaAs distributed Bragg reflector (DBR) with the observation of Fabry–Pérot (FP) modes under optical excitation. The process for obtaining this monolithic integration of [111]-oriented nanowires on (001)-oriented DBR is illustrated in Scheme 1.

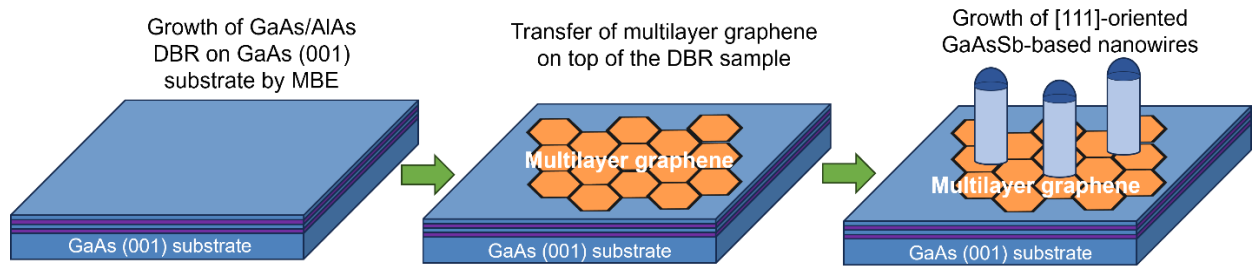

**Scheme1.** Illustration of the process for monolithic integration of vertical [111]-oriented GaAsSb-based nanowires on (001) GaAs/AlAs DBR using Scotch exfoliation transfer method of multilayer graphene intermediate layer and MBE.

To firstly explore the nucleation and growth of GaAsSb nanowires on graphitic substrate, kish graphite was indium-bonded to a Si(001) wafer as substrate for the growth of GaAsSb nanowires in a Veeco Gen 930 MBE system. Before growth, the substrates were outgassed in the buffer chamber at 550 °C until the chamber pressure reached ~ $2\times10^{-8}$ Torr. In this work, the group-III fluxes (Ga and Al) are expressed as equivalent growth rates (monolayers per second, ML/s) for GaAs and AlAs thin films on a GaAs(001) substrate as determined by reflection high-energy electron diffraction (RHEED). The group-V fluxes ($As_2$ and $Sb_2$) were measured in terms of beam equivalent pressure (BEP) by an ion gauge. For the ALE growth study, we used an Al equivalent growth rate of 0.3 ML/s and $As_2$ and $Sb_2$ BEP of $2.5\times10^{-7}$ and $7.5\times10^{-7}$ Torr, respectively. These ALE experiments were performed at 140 °C or 150 °C, as measured by a thermocouple, to minimize the aggregation of adatoms. One cycle of the ALE process contains two individual deposition steps for either group-III or -V species, which were controlled by opening the correlated shutter(s) for 3 s. The nanowire growth experiments were carried out at 630 °C, as measured by a pyrometer. The thin-film DBR structure was grown at 585 °C on a GaAs(001) wafer, comprising 20 pairs of AlAs/GaAs layers with individual layer thicknesses of 80.93 nm and 65.22 nm, respectively.

Although direct epitaxial growth of GaAs nanowires on kish graphite and epitaxial graphene on SiC has been realized by the self-catalyzed growth method using low-temperature nucleation and high-temperature growth scheme[9], there are yet few reports of GaAsSb nanowire growth on graphitic substrates using the same strategy. We here first employed a similar two-temperature scheme as used for GaAs nanowire growth on graphene.[9] A Ga predeposition step was performed at 630 C° by first supplying a Ga flux of 0.7 ML/s for 10 s. Then, the Ga shutter was closed and the substrate temperature was ramped down to 540 C° for the nucleation step, performed by supplying an $Sb_2$ flux of $8\times10^{-7}$ Torr for 10 s. Keeping the $Sb_2$ flux, an $As_2$ flux of $2.5\times10^{-6}$ Torr was then supplied together with the $Sb_2$ flux for the next 40 s (without any Ga flux). For the GaAsSb nanowire growth, the substrate temperature was ramped up to 630 °C for 10 min growth using Ga, $Sb_2$ and $As_2$ without changing the fluxes, which for our purpose is an optimal growth condition for this GaAsSb nanowire growth[29].

Before the growth study, the surface morphology of kish graphite was characterized by atomic force microscopy (AFM) using the tapping mode of a Bruker Dimension Icon system. The height profile micrograph and its 3D projection are shown in Fig. 1a and b, respectively, demonstrating an atomically flat surface, ideal for the epitaxial experiments. A scanning electron microscopy (SEM) image of the grown sample was taken with a Zeiss Supra 55 with in-lens secondary electron detector and is shown in Fig. 1c. It reveals large Ga droplet formation without any in-plane or vertical nanowire growth on the graphite surface, indicating an absence of effective nucleation of GaAsSb nanowires on the kish graphite surface.

To understand the nanowire nucleation mechanism on graphitic surfaces, especially the Sb influence, we look at the nucleation model of nanowire growth[34]. The formation enthalpy of III-V nuclei on a graphitic surface can be expressed as

$$\Delta G = -A \cdot h \cdot \Delta\mu + P \cdot h \cdot \gamma_{lL} + A \cdot (\gamma_{NL} - \gamma_{GL} + \gamma_{GN}), \qquad (1)$$

where *A, h* and *P* are the upper surface area, height and perimeter, respectively, of the III-V nuclei on graphite, $\Delta\mu$ is the supersaturation of the group-V elements in the Ga droplet. $\gamma_{lL}$ and $\gamma_{NL}$ are the energies per unit area of the lateral and top interface, respectively, between nucleus and liquid. $\gamma_{GL}$ and $\gamma_{GN}$ are the energies per unit area of the graphite-liquid and graphite-nucleus interfaces, respectively. The driving force for the nanowire growth is the supersaturation $\Delta\mu$. Considering the surface energy of liquid Ga of 0.7 J/m$^2$ and of GaAs(111)B of 1.3 J/m$^2$,[35,36] $\gamma_{NL}$ and $\gamma_{GL}$ are estimated to be around 0.9 J/m$^2$ and 0.4 J/m$^2$ due to the low energy of the graphitic-vapor interface (0.06 J/m$^2$).[37,22] Due to $\gamma_{GL}$ being much smaller compared to $\gamma_{NL}$, the surface energy term is a significant barrier for III-V nucleation on graphitic surfaces, which requires higher supersaturation for a negative formation enthalpy, as seen from equation 1. It has been reported that the incorporation of Sb will decrease the supersaturation of self-catalyzed GaAs nanowires due to a compositional influence, suggesting a smaller $\Delta\mu$ in equation 1 by growing GaAsSb nanowires instead of GaAs ones.[29] Apart from its compositional influence, Sb can also contribute to a reduction in the energy per unit area at the interface between graphite and liquid Ga, $\gamma_{GL}$, through its surfactant effect.[30,38] Since the supersaturation is the driving force to overcome the nucleation barrier and form nuclei of size suitable for further growth on graphite, the Sb-induced decrease of the supersaturation and the surfactant effect make the growth of GaAsSb nanowires more challenging than that of GaAs nanowires on graphitic surfaces, which can well explain the droplets left on the graphitic surface without any effective crystal nucleation shown in Fig. 1c. We note that the nucleation and growth of high-Sb content GaAsSb nanowires is also very challenging on Si[25].

Although the self-catalyzed GaAs nanowires have enough supersaturation for the nuclei formation on graphitic surfaces, the vertical nanowire density is still very low and the surface parasitic growth is eminent.[9,22] This non-vertical nanowire growth should come from the low surface energy of the chemically inert sp$^2$ bonded graphitic surfaces, which results in a Volmer–Weber island growth mode.[22] Since the catalyst-assisted vertical nanowire growth proceeds in a layer-by-layer manner[39,40], it is essential to increase the surface energy of the graphitic surfaces, and enhance the binding between the III-V materials and the graphitic surface to facilitate the nanowire growth on graphitic surfaces in the regime of layer-by-layer mode.[29] To realize vertical growth of GaAsSb nanowires, the formation energy with respect to the nucleus-graphite interface, $\gamma_{GN}$, in equation 1 was investigated. When $\gamma_{GN}$ is large, it will lead to either no nucleation due to insufficient supersaturation or in-plane nanowire growth due to the difficulty of a complete monolayer formation in the catalyst[22]. To reduce $\gamma_{GN}$, increasing the surface energy of kish graphite is one possible solution. It has been reported that an $O_2$ plasma treatment can promote InAs nanowires growth on graphitic surfaces.[41] Here, we use a UV-ozone and hydrogenation process, $O_2$ flushed chamber at ambient pressure, to pretreat the graphitic surface. This process generates ions of lower kinetic energy in a much softer (more gentle) and controllable way than using $O_2$ plasma. The soft radicals of $O_3$ can create monolayer-thick step edges of about 330 pm height, as shown in Fig. 1d and e, and these step edges contain high-energy dangling bonds which increase the surface energy of kish graphite. Growth on such kish graphite with introduced step-edges following a similar two-step growth recipe as described above, except that the nucleation temperature was reduced to 250 °C, resulted in the morphology as shown in Fig. 1f. It shows a high-density nanowire growth, correlating well with the analysis of the nucleation model of equation (1) presented above. We note a recent study shows the GaAsSb nanowire growth on a single layer graphene/$SiO_2$/Si(111) substrate after an $O_2$ plasma treatment.[42]

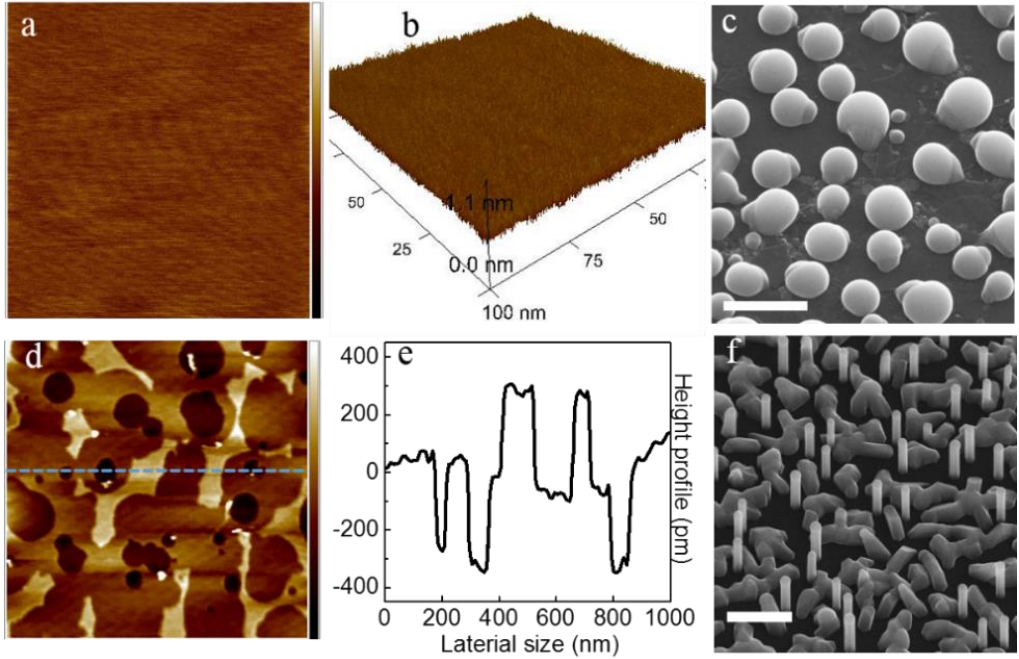

**Figure 1.** (a) Top-view and (b) tilted-view AFM image of pristine kish graphite surface before nanowire growth. (c) GaAsSb nanowire growth attempt on kish graphite without ozone pretreatment using a two-step growth method, with nucleation at 540 °C and growth at 630 °C. (d) Top-view AFM image of kish graphite after ozone treatment for 9 min. (e) Height profile of ozone treated kish graphite along the blue-dash line in (d). (f) GaAsSb nanowires grown on kish graphite with ozone pretreatment, with nucleation at 250 °C and growth at 630 °C. The lateral dimensions of the AFM images in (a) and (d) are 100 nm and 1 μm with the range of the heights as shown in (b) and (e), respectively. The scale bars in (c) and (f) are 500 nm and 1 μm, respectively.

Although we have successfully grown GaAsSb nanowires on graphitic surfaces that have undergone soft UV-ozone and hydrogenation treatment, the damage of the kish graphite may not be desired for applications when single or few layer graphene is needed, e.g., functioning as a transparent contact electrode.[43] To explore the effective nucleation condition on pristine graphitic surfaces, we used ALE to study the interaction between graphite and the III-V materials, by alternating the deposition sequence of group-III element Al and group–V elements $Sb_2$ and $As_2$, respectively. Fig. 2a depicts a schematic of the MBE deposition of Al, As and Sb materials onto graphitic substrates from separate crucibles under high vacuum conditions. The grey triangles represent AlAsSb nuclei with three-fold symmetry that are well aligned with each other and the graphitic substrate. In the SEM image of Fig. 2b, triangular AlAsSb crystallites on kish graphite can be clearly identified, resulting from three cycles at 150 °C initiated with an Al deposition in the ALE process. According to the Wulff construction with minimized surface free energy, the three edge facets of the triangle should belong to either the {11-2} or the {1-10} family of planes.[44,45] This symmetry suggests a [111] growth direction, and the next layer has truncated triangle or even a hexagonal 6-fold symmetry. We further reduced the growth temperature by 10 °C to lower the diffusion difference among Al, Sb and As, and nano-crystallites of other three-fold symmetry were indeed observed, as shown in Fig. 2c. In the inset of Fig. 2c, two different three-fold symmetries are marked in red and black dashed lines. From the above observation with aligned crystals of three-fold symmetry, it is concluded that the triangles formed the same epitaxy relation with the central hexagonal nucleus in Fig. 2b and is consistent with recent growth and characterization of qvdW epitaxy[10]. Since the ALE process has separate group-III and -V deposition steps, the edge facets can easily be passivated/saturated by either the group-III or -V element, leading to a faster growth rate in three edge facets instead of six facets and thereby forming the observed triangular shape instead of a hexagonal morphology.[46] After confirming the successful nucleation of AlAsSb nanostructures on graphite by using an ALE process initiated with Al deposition, it is interesting to compare the nucleation with an ALE process initiated by As and Sb deposition, which is shown in Fig. 2d. It is clear that the nanocrystals formed by the ALE process initiated by an As and Sb deposition do not show any pronounced symmetry or well-defined structure. This implies that these nanostructures are either amorphous or highly defective. Although the ALE processes have similar growth temperature and group-III and -V fluxes, the distinct morphologies suggest that the initial interfacial interaction between graphite and III-V materials is crucial for the successful formation of III-V/graphite hybrid structures with high crystalline quality.

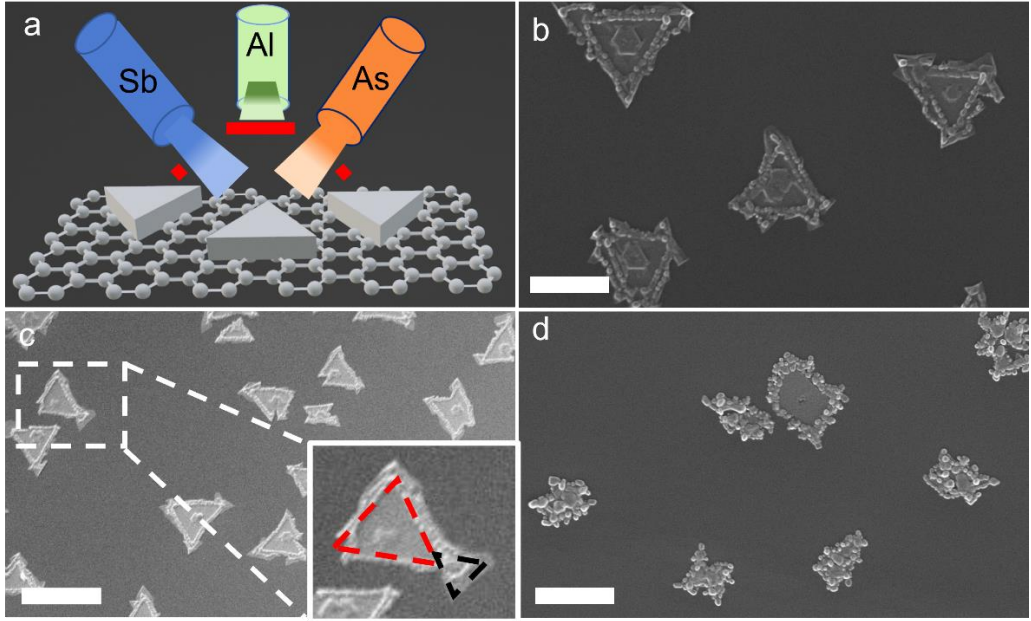

**Figure 2.** (a) Schematic of ALE process with separate deposition steps of group-III and -V elements. SEM images of ALE process with three repeated Al/(As, Sb) depositions (b) at 150 °C and (c) at 140 °C, and (d) (As, Sb)/Al depositions at 150 °C. The scale bars in (b), (c) and (d) are 500 nm, 1 μm and 500 nm, respectively.

To successfully grow high-density GaAsSb nanowires, we supplied 0.05 ML/s of Al for 1 s, and then $As_2$ and $Sb_2$ at $2.5\times10^{-6}$ and $6\times10^{-7}$ Torr, respectively, for 1 min as a nucleation step at 630 °C. This process of high-temperature nucleation minimizes the unwanted 2D parasitic growth and enables the growth of long nanowires with sufficient thickness for use in photonic applications. Then the GaAsSb nanowire growth was carried out by opening the Ga (flux of 0.7 ML/s), As and Sb shutters simultaneously for 5 min. As can be seen from the SEM image in Fig. 3a, a high-density GaAsSb nanowire growth was achieved on graphite without any pretreatment. Although the growth of GaAsSb is more challenging than GaAs, considering the effects of the introduction of Sb, using AlAsSb nucleus results in an even higher density than previous reports on the growth of GaAs nanowires.[9,22] To characterize the epitaxial relation between the GaAsSb nanowires and the graphite substrate, we performed cross-section scanning transmission electron microscopy (STEM) characterization. Bright-field STEM imaging, Fig. 3b, clearly shows a nanowire with the growth direction vertical to the graphite substrate. The high-resolution and high-angle annular dark-field STEM micrograph from the red rectangle in Fig. 3b at the graphite/nanowire interface, shown in Fig. 3c, confirms that the nanowire growth is along the [111] direction, that the lower part is Al rich ($Z = 13$) giving a lower intensity compared to the upper part which is Ga-rich ($Z= 31$) and that the nanowire is predominantly of ZB crystal phase with few stacking faults where the composition is changing from Al- to Ga-rich. The upper GaAsSb part has a stable epitaxial growth condition with a very low stacking fault/twin density, in line with what is commonly observed for GaAsSb nanowire growth. The Sb composition of the nanowires with AlAsSb buffer was measured by energy-dispersive X-ray spectroscopy, showing an Sb molar fraction of about 35 %, i.e., $GaAs_{0.65}Sb_{0.35}$, for the nanowires grown on kish graphite.

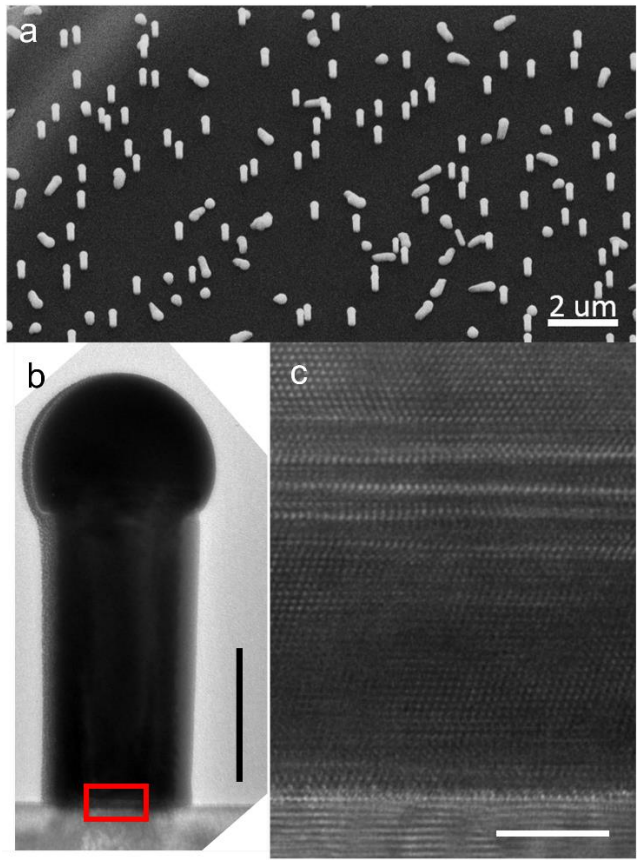

**Figure 3.** (a) SEM image of high-density GaAsSb nanowires grown on graphite using AlAsSb for the nucleation. (b) Cross-section bright-field STEM of a GaAsSb nanowire. (c) Lattice imaging-high-angle annular dark-field STEM from graphite/nanowire interface of the rectangle area marked in red in (b). The scale bars for the TEM images are 200 nm and 5 nm for (b) and (c), respectively.

As previously reported, the grown III-V materials and LED devices can be easily lifted off for flexible application by using graphene as an intermediate layer[5]. Here, we utilize another very important advantage of graphene/graphite, i.e., the flexibility for integration on an arbitrary platform by the exfoliation and transfer method, to integrate as-grown III-V nanowires with a DBR. Although the III-V nanowires grow in the [111] directions and the GaAs/AlAs DBR has best growth conditions in the [001] direction, exfoliated graphitic flakes can make a perfect transition as a buffer to initiate the [111] nanowire growth on a (001) GaAs substrate, as illustrated in Scheme 1. In this study, twenty periods of a GaAs/AlAs quarter-wavelength DBR (with central high reflectivity at 930 nm) was grown on a GaAs(001) substrate, and exfoliated graphitic flakes were transferred onto the top of the DBR structure using the Scotch tape method[12]. Transferred graphite flakes on top of the DBR can be seen in the optical image in Fig. 4a. GaAsSb-based nanowire superlattice structures that have previously shown excellent room-temperature lasing behavior[28] were grown on the DBR with graphitic flakes using the optimized AlAsSb nucleation as described above. Fig. 4b is an SEM image of a single nanowire grown on top of the thin graphitic flake with a size of around 5 μm located on the top left of the marker letter "C" in Fig. 4a. The inset of Fig. 4b shows a larger overview of the single vertical nanowire grown on the exact graphitic flake in Fig. 4a and it confirms that a GaAsSb thin film structure grew directly on the DBR outside the graphitic flakes and vertical nanowires grew selectively on top of the DBR with graphitic flakes. The DBR reflectivity was measured by a Filmetrics optical reflectometer. As a near-perfect high reflectivity plateau with close to 100 % reflectivity and beyond the resolution of the reflectometer was observed from 890 nm to 960 nm, the measured DBR reflectivity was normalized in Fig. 4c. To acquire the optical signal from the single as-grown nanowire that interacts with the bottom DBR as shown in Fig. 4b, the absorption was maximized by tilting the nanowire/DBR sample at ~ 45 ° with respect to the excitation beam using femtosecond pulses at 800 nm from a Ti:sapphire laser. The schematic geometry of the experimental setup for the optical pumping experiments is shown in Fig. 4d. The photoluminescence (PL) emission from the nanowire/DBR sample was collected by the same objective and guided through several beam splitters to a spectrometer. PL spectra at different excitation powers are plotted in Fig. 4c together with the DBR reflectivity. Unlike typical PL from nanowires on substrates without optical feedback, periodic modulation of the emitted PL intensity can be clearly observed from 900 nm to 980 nm. The PL peaks indicating the presence of Fabry–Pérot (FP) modes due to the as-grown nanowire cavity defined by a DBR bottom reflector and a high-index contrast top nanowire end facet are marked with grey dashed lines in Fig. 4c. Considering the length (~ 6.6 μm) and thickness (~ 400 nm) of the nanowire in Fig. 4b, the inter-modal spacing of ~ 17.4 meV agrees well to the inverse relation between the length of the FP cavity and the energy spacing between two FP modes.[47] We would like to mention that the energy spacing of two adjacent FP modes in our GaAsSb nanowires is larger compared to the ones mentioned in ref. 28 and 47. This is primarily attributed to the shorter length of the nanowire cavity in our study. The FP modes overlap well with the high-reflectivity plateau of the DBR, which further confirms that there is substantial optical feedback to the nanowire from the DBR. The intensity of the modes increases when the excitation power is increased from 0.5 to 1.5 mW, and there is some gain when the excitation is increased. However, the gain saturates when the power is increased beyond 1.5 mW. For the PL spectrum without feedback from the DBR substrate, we refer to the sample grown on Si, whose PL was shown in Fig. 1i in Ref. 28. The growth procedure for those nanowires with axial heterostructures was identical, except for the nucleation step. No observable feedback was present for those nanowires grown directly on Si(111). We note that the total nanowire length used in this work is shorter than in ref. 28. This is due to the significant thin-film growth on the DBR outside the graphitic flakes as can be seen in Fig. 4(b), which consumes significant amount of Ga and slows down the axial nanowire growth[31]. Further research of positioned nanowire growth on graphene will be highly needed to achieve optimized geometry of nanowire cavity towards high-quality polarization measurements[48] and lasing operation at room temperature[28].

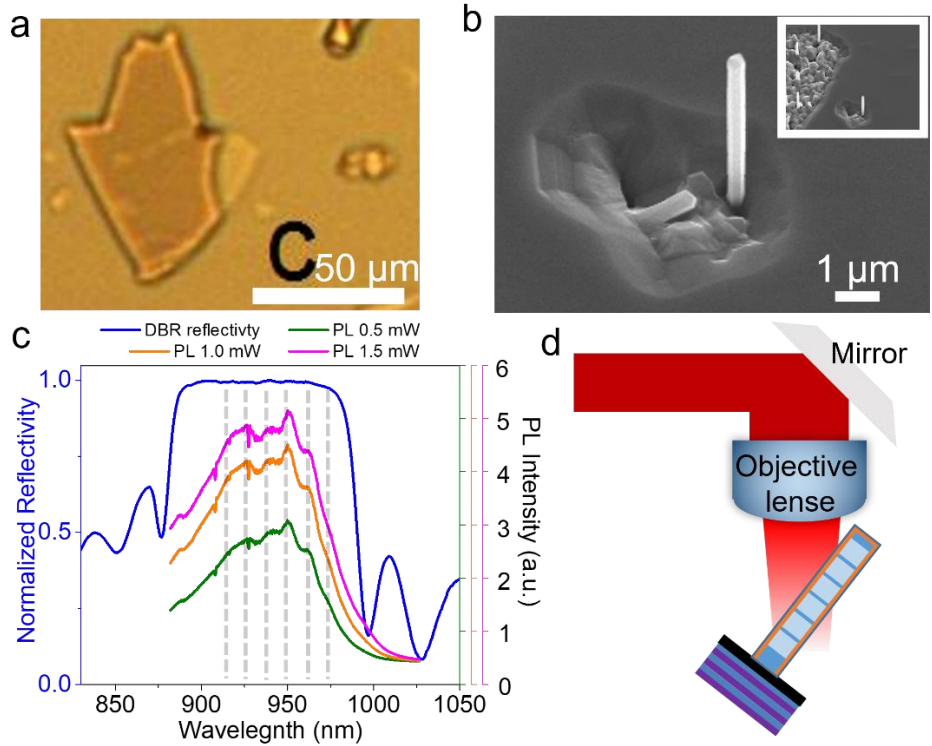

**Figure 4.** (a) Optical micrograph image of exfoliated kish graphite on a twenty period AlAs/GaAs DBR grown on a (001) GaAs substrate. (b) Tilted-view SEM image of GaAsSb/GaAs superlattice nanowire on graphite/DBR shown on the top left of the letter "C" in (a). A zoom-out image is shown in the inset. (c) Measured reflectivity of the AlAs/GaAs DBR and the PL signal from the nanowire in (b) at optical excitation powers of 0.5, 1 and 1.5 mW. (d) Schematic of the geometry for the optical PL measurements of the GaAsSb/GaAs nanowire on the DBR.

In summary, it is shown that the growth of self-catalyzed GaAsSb nanowires on graphitic substrates can be realized by creating step edges of monolayer-thickness on a kish graphite surface before growth or by using an AlAsSb nucleation buffer. Atomic layer deposition experiments confirm that the bonding between Al and graphitic surface is preferential for nucleation compared to the bonding between the graphitic surface and the group-V elements. Using a Scotch tape method, graphitic flakes were transferred onto foreign substrates to realize monolithic integration of [111]-oriented GaAsSb-based superlattice nanowires on a high-quality quarter-wavelength (001) AlAs/GaAs DBR. By performing optical pumping experiments on an as-grown nanowire, Fabry–Pérot resonance modes were observed and the energy spacing between two FP resonances is well correlated to the cavity length, indicating a successful integration of the nanowire active medium onto the external DBR substrate. These results highlight the potential for seamlessly integrating highly efficient photonic nanowire devices onto various photonic platforms using a thin graphitic layer. For instance, epitaxy-grown III-V nanowires on amorphous DBRs could be employed for vertical emitting LEDs and laser devices.

AUTHOR INFORMATION

**Corresponding Author**

* Correspondence should be addressed to: bjorn.fimland@ntnu.no

ACKNOWLEDGMENT

The Research Council of Norway is acknowledged for the support for NANO2021 grant number 239206, and to the Norwegian Micro- and Nano-Fabrication Facility, NorFab, grant number 295864 and NORTEM, grant number 197405.

Table of content (TOC)

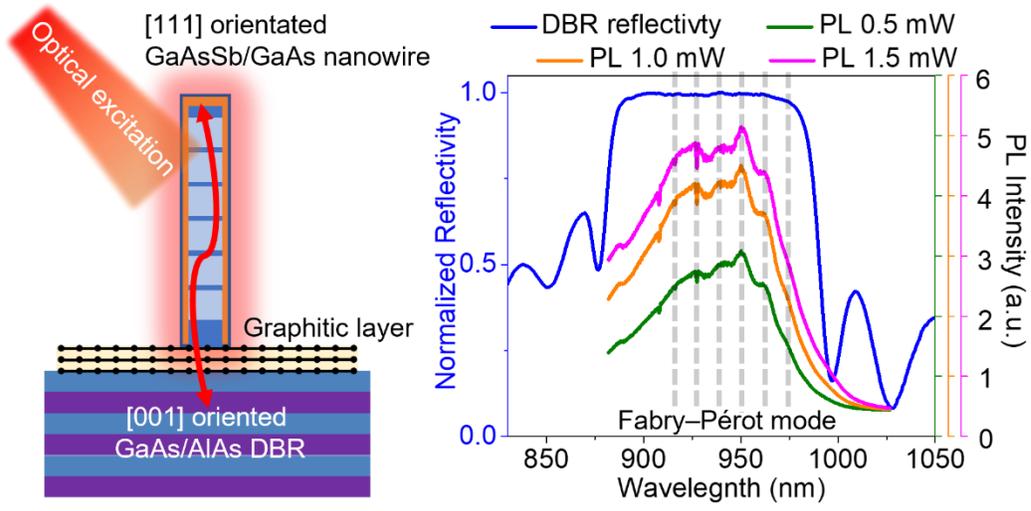